\documentclass[12pt]{iopart}
\usepackage{graphicx,color}
\usepackage{amssymb}
\usepackage{graphicx}
\usepackage{epsf}
\usepackage{bm}

\begin{document}

\def\um{\mu\mbox{m}}
\def\sec{\mbox{s}}
\def\psec{\mbox{ps}}
\def\fsec{\mbox{fs}}
\def\Wcm2{\mbox{W cm}^{-2}}
\def\Wcmum2{\mbox{W cm}^{-2}\mu\mbox{m}^{2}}
\def\Vcm{\mbox{V cm}^{-1}}
\def\Vm{\mbox{V m}^{-1}}
\def\cm3{\mbox{cm}^{-3}}

\title{Dynamics of charge-displacement channeling in intense laser-plasma
interactions}

\author{S.~Kar\footnote{Electronic address: s.kar@qub.ac.uk}, M.~Borghesi, C.~A.~Cecchetti, L.~Romagnani}
\address{School of Mathematics and Physics, the Queen's University of
Belfast, Belfast BT7 1NN, UK}
\author{F.~Ceccherini, T.~V.~Liseykina\footnote{On leave from the Institute for Computational Technologies, SD-RAS, Novosibirsk, Russia}, A.~Macchi\footnote{also
affiliated to polyLab, CNR-INFM, Pisa, Italy}}
\address{Dipartimento di Fisica ``E. Fermi'', Universit\`a di
Pisa, Pisa, Italy}
\author{R.~Jung, J.~Osterholz, O.~Willi}
\address{Institut f\"ur Laser-und Plasmaphysik,
Heinrich-Heine-Universit\"at, D\"usseldorf, Germany}
\author{L.~A.~Gizzi}
\address{Intense Laser Irradiation Laboratory, IPCF-CNR, Pisa,
Italy}
\author{A.~Schiavi}
\address{Dipartimento di Energetica, Universit\`a di Roma 1 ``La
Sapienza'', Roma, Italy}
\author{M.~Galimberti, R.~Heathcote}
\address{Central Laser Facility, Rutherford Appleton Laboratory,
Chilton, OX11 0QX, UK}

\begin{abstract}

The dynamics of transient electric fields generated by the
interaction of high intensity laser pulses with underdense plasmas
has been studied experimentally with the proton projection imaging
technique. The formation of a charged channel, the propagation of
its front edge and the late electric field evolution have been
characterised with high temporal and spatial resolution.
Particle-in-cell simulations and an electrostatic, ponderomotive
model reproduce the experimental features and trace them back to the
ponderomotive expulsion of electrons and the subsequent ion
acceleration.

\end{abstract}


\maketitle

\section{Introduction}

The study of the propagation of intense laser pulses in underdense
plasmas is relevant to several highly advanced applications,
including electron \cite{electron-acceleration} and ion acceleration
\cite{ion-acceleration,ion-acceleration-2}, development of X- and
$\gamma$-ray sources \cite{xray}, and fusion neutron production
\cite{Fritzler_PRL_02}. It is also of fundamental interest, due to
the variety of relativistic and nonlinear phenomena which arise in
the laser-plasma interaction \cite{Bulanov_RPP_01}. Among these,
self-focusing and self-channeling of the laser pulse arise in this
regime from the intensity dependence of the relativistic index of
refraction \cite{Sun_PF_87,Mori_PRL_88}.

Strong space charge electric fields are generated during the early
stage of the propagation of a superintense laser pulse through an
underdense plasma as the ponderomotive force acts on electrons,
pushing them away from the axis. Thus, for a transient stage the
pulse may propagate self-guided in a charged channel
\cite{Borisov_PRL_92}, while the space-charge field in turn drags
and accelerates the ions to MeV energies\cite{ion-acceleration-2}.
So far, experiments have provided evidence of channel formation and
explosion using optical
diagnostics~\cite{Borisov_PRL_92,Monot_PRL_95,Borghesi_PRL_97,Krushelnick_PRL_97,Sarkisov_JETP_97,Sarkisov_PRE_99}.
or by detecting radially accelerated
ions~\cite{ion-acceleration,ion-acceleration-2,Fritzler_PRL_02,Sarkisov_PRE_99},
while a direct detection of the space-charge fields has not been
obtained yet. The development of the the proton projection imaging (PPI)
technique~\cite{Borghesi_RSI_03} has provided a very powerful tool
to explore the fast dynamics of plasma phenomena via the detection
of the associated transient electric field structures. The technique
is based on the use of laser-accelerated multi-MeV protons
(\cite{fuchsNP06,robsonNP06} and references therein) as a charged
probe beam of transient electromagnetic fields in plasmas, a
possibility allowed by the low emittance and high laminarity of the
proton source~\cite{borghesiPRL04,cowanPRL04}, as well as by its
ultra-short duration and the straightforward synchronization with an
interaction laser pulse. The experimental PPI implementation takes
advantage from the broad energy spectrum of protons, since in a
time-of-flight arrangement protons of different energy will probe
the plasma at different times, and thus an energy-resolved
monitoring of the proton probe profile allows to obtain single-shot,
multi-frame temporal scans of the
interaction~\cite{Borghesi_RSI_03}. PPI and the related "proton
deflectometry" technique permit to gather spatial and temporal maps
of the electric fields in the plasma, and therefore have proven to
be an unique tool to explore the picosecond dynamics of laser-plasma
phenomena~\cite{borghesiPRL02,borghesiPRL05,romagnaniPRL05} via the
associated space-charge fields.

In this article, we report on an experiment using the PPI technique
to study of the formation and subsequent evolution of a
charge-displacement channel in an underdense plasma. These
investigations have led to the first direct experimental detection
of the transient electric fields in the channel, providing an
insight of the fundamental physical processes involved. The
comparison of the experimental data with two-dimensional (2D)
electromagnetic (EM) particle-in-cell (PIC) simulations and a simple
one-dimensional (1D) electrostatic (ES) PIC model allows to
characterize in detail the electric field dynamics at different
stages of its evolution.

\section{Experimental setup}

\begin{figure}[h]
\begin{center}
\includegraphics[angle=0,width=0.75\textwidth]{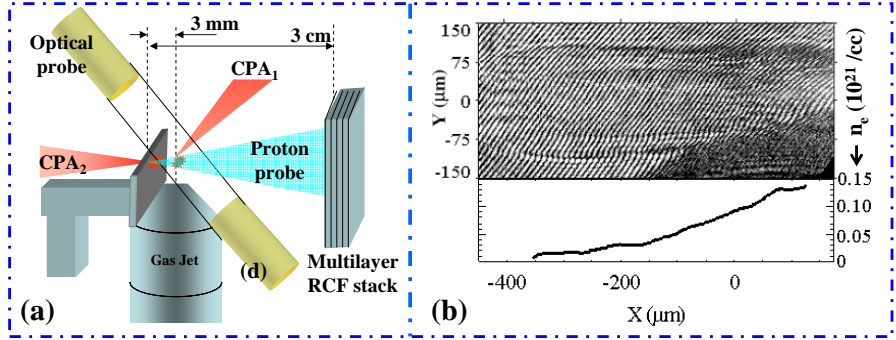}
\end{center}
\caption{(color) \textbf{(a)} Schematic of the experimental setup.
{\textbf{(b)} Top: Interferogram of the plasma at 25~ps before the
arrival of $\mbox{CPA}_{1}$ at its focal plane $x=0$. The
$\mbox{CPA}_{1}$ peak intensity was $1.5\times 10^{19}~\Wcm2$.
Bottom: the corresponding electron density profile along the $y=0$
axis.}\label{expt_setup}}
\end{figure}

The experiment was carried out at the Rutherford Appleton
Laboratory, employing the VULCAN Nd-Glass laser system
\cite{Danson_JMO_98}, providing two Chirped Pulse Amplified (CPA)
pulses, with $1.054~\um$ wavelength, synchronized with picosecond
precision. Each of the beams delivered approximately 30~J on target
in 1.3~ps (FWHM) duration. By using $f/6$ off-axis parabolas, the
beams were focused to spots of $10~\um$ (FWHM) achieving peak
intensities up to $3 \times 10^{19}~\Wcm2$. The short pulses were
preceded by an Amplified Spontaneous Emission (ASE) pedestal of
300~ps duration and contrast ratio of $\sim 10^{6}$
\cite{Gregori_PC_06}. One of the beams ($\mbox{CPA}_{1}$) was
directed to propagate through He gas from a supersonic nozzle,
having a 2~mm aperture, driven at 50 bar pressure. The interaction
was transversely probed by the proton beam produced from the
interaction of the second CPA beam ($\mbox{CPA}_{2}$) with a flat
foil (a $10~\um$ thick Au foil was typically used), under the point
projection imaging scheme \cite{Borghesi_RSI_03}. The schematic of
the experimental setup is shown in Fig.\ref{expt_setup}(a). Due to
the Bragg peak energy deposition properties of the protons, the use
of multilayered stacks of Radiochromic film (RCF) detector permits
energy-resolved monitoring of the proton probe profile, as each
layer will primarily detect protons within a given energy range.
This allows to obtain single-shot, multi-frame temporal scans of the
interaction in a time-of-flight arrangement \cite{Borghesi_RSI_03}.
The spatial and temporal resolution of each frame were of the order
of a few $\psec$, and of a few microns, respectively, while the
magnification was 11.

The interaction region was also diagnosed by Nomarsky
interferometry, employing a frequency doubled CPA pulse of low
energy. The reconstructed electron density profile along the
propagation axis before the high-intensity interaction [see
Fig.~\ref{expt_setup}(b)], broadly consistent with the neutral
density profile of the gas jet \cite{Jung_CLF_2005} suggests
complete ionisation of the gas by the ASE prepulse.

\section{Experimental Results}

\begin{figure}[h]
\begin{center}
\includegraphics[angle=0,width=0.7\textwidth]{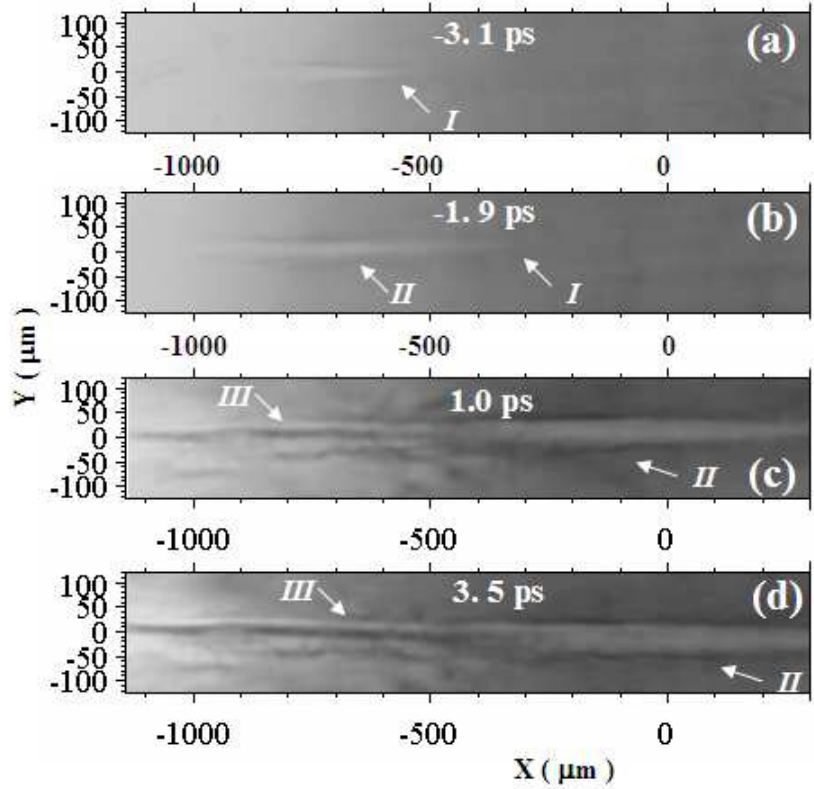}
\end{center}
\caption{Proton projection images of the interaction region at
different times obtained in two different laser shots. The $x$ and
$y$ coordinates refer to the object (interaction) plane, which
intersects the probe axis at $(x,y)=(0,0)$. The images {(a)} and
{(b)}  have been obtained from the same shot at an intensity
$I=4.0\times10^{18}~\Wcm2$, while {(c)} and {(d)} correspond to a
shot with  $I=1.5\times10^{19}~\Wcm2$. The signal in the frames is
mainly due to protons [of energies E = 13 MeV, 12.5 MeV, 13 MeV and
12 MeV in (a)-(d) respectively] reaching the Bragg peak within their
active layers. The time labels give the probing time of the protons
propagating along the probe axis [$t_0(E)$], relative to the time of
arrival of the peak of the interaction pulse at the plane $x=0~\um$.
White (dark) regions correspond to lower (higher) proton flux than
the background. The labels indicate the most prominent features: I)
the bullet-shaped leading edge and II) the central region of the
'white', positively charged channel; III) the 'black' line along the
axis, indicating a region of field inversion inside the channel.
\label{channel}}
\end{figure}

Fig.~\ref{channel}(a), (b) and (c) show three sequential
proton-projection images of the interaction region. The laser pulse
propagates from left to right. A 'white' channel with 'dark'
boundaries is visible at early times [Fig.\ref{channel} (a) and (b)].
The leading, 'bullet' shaped edge of the channel, indicated by the label I in
Fig.\ref{channel} (a) and (b), is seen moving along the laser axis.
In the trail of the channel the proton flux
distribution changes qualitatively [Fig.~\ref{channel}(c) and (d)],
showing a dark line along the axis (indicated by the label III),
which is observed up to tens of
ps after the transit of the peak of the pulse. The 'white' channel
reveals the presence of a positively charged region around the
laser axis, where the electric field points outwards.
This can be interpreted as the result of the expulsion of electrons
from the central region. The central dark line observed at later
times in the channel suggests that at this stage the radial electric
field must change its sign, i.e. point inwards, at some radial
position. As discussed below, this field inversion is related to the
effects of ion motion.

\begin{figure}[h]
\begin{center}
\includegraphics[angle=0,width=0.4\textwidth]{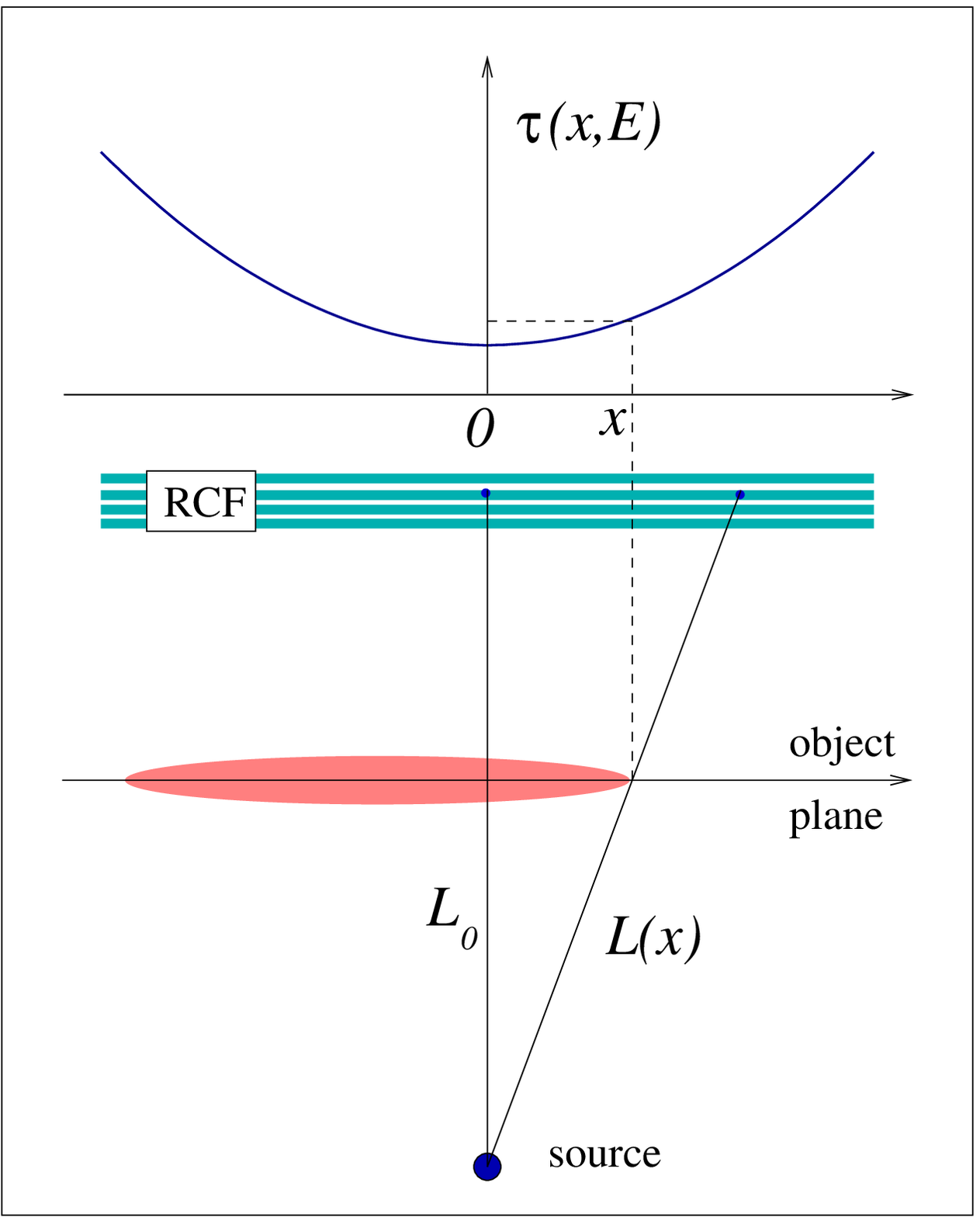}
\end{center}
\caption{Schematic showing the evaluation of the channel front
speed. Protons of energy $E$ from the point-like source cross the
channel axis at a distance $x$ from the centre of the object plane
at the time $\tau(x,E) = t_0(E) +[L(x)-L_0]/v_p$, where $L_0$ is the
distance between the source and the object plane,
$v_p=\sqrt{2E/m_p}$ is the proton velocity, $L(x)=\sqrt{x^2+L_0^2}$,
and $t_0(E)$ is the time at which the protons of energy $E$
directed perpendicularly to the object plane cross the latter,
relative to the instant at which the laser pulse peak reaches
the focal plane ($x=0$).
\label{fig:probingtime}}
\end{figure}

Due to multi-frame capability of the PPI technique with ps temporal
resolution, it has been possible to estimate the propagation
velocity $v$ of the channel front. Critically, one has to take into
account that, due to the divergence of the probe beam, the probing
time varies along the pulse propagation axis (see
Fig.~\ref{fig:probingtime}) as $\tau(x,E) \simeq t_0(E)
+\tau_0(E)(\sqrt{1+x^2/L_0^2}-1)$ where $L_0 \simeq 0.3~\mbox{cm}$
is the distance between the plane and the proton source and
$\tau_0=L_0/\sqrt{2E/m_p} \simeq 220~\psec/\sqrt{E/{\mbox{MeV}}}$ is
the proton time of flight from the source to the center of the
object plane (see Fig.\ref{fig:probingtime}).
The reference time $t_0(E)$ is relative to the instant at which the
laser pulse peak crosses the focal plane $x=0$. We divide the
displacement of the tip of the channel leading front
from frame (a) to (b), $\Delta X
=X_{b}-X_{a} \simeq (-300+500)~\um=200~\um$ by the difference
$\Delta\tau$ in the corresponding probing times
$\Delta\tau=\tau(X_b,E_b)-\tau(X_a,E_a)\simeq 0.66~\psec$ to obtain
$v=\Delta X/\Delta\tau \simeq 3\times 10^8~\mbox{m/s}$. Based on the
nominal reference time we estimate the peak of the pulse to be
approximately $0.6~\psec$ behind the tip of the channel front.

\section{Data analysis by numerical simulations}

In order to unfold the physical mechanisms associated with the
dynamics of the charged channel, a 1D ES PIC model in cylindrical
geometry was employed, in which the laser action is modeled solely
via the ponderomotive force (PF) of a non-evolving laser pulse. A
similar approach has been previously used by other authors
\cite{Krushelnick_PRL_97,Sarkisov_PRE_99}. The code solves the
equation of motion for plasma particles along the radial direction,
taking into account the ES field obtained from Poisson's equation
and the PF acting on the electrons -
$F_r=-m_ec^2\partial_r[1+a^2(r,t)/2]^{1/2}$~\cite{bauerPRL95}. Here
$a(r,t)=a_0e^{-r^2/2r_0^2}f(t)$ and $f(t)$ defines the temporal
envelope of the laser pulse. For the latter, a '$\sin^2$' profile
was used (the use of a Gaussian profile did not yield significant
differences).

\begin{figure}[h]
\begin{center}
\includegraphics[angle=0,width=0.7\textwidth]{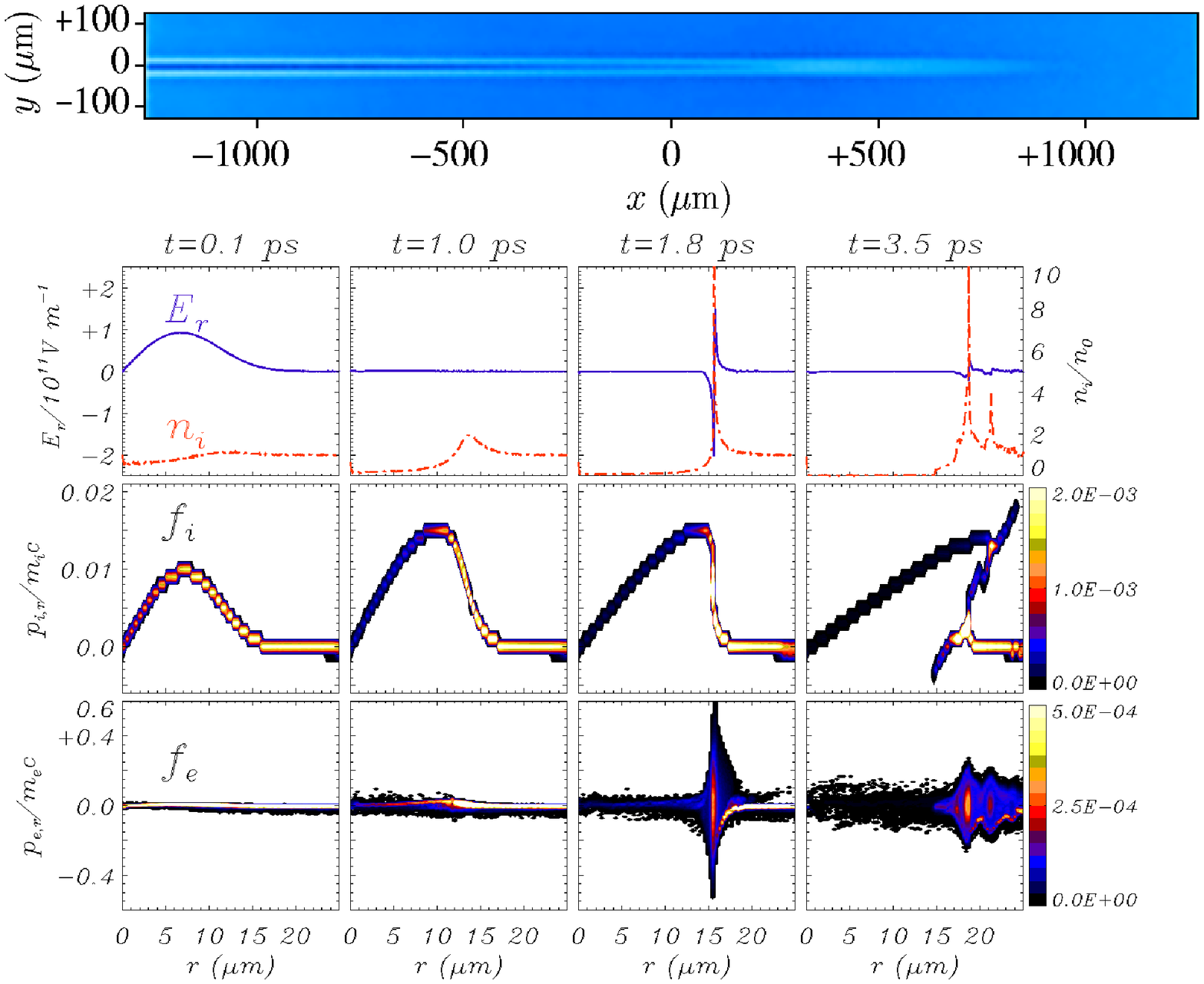}
\end{center}
\caption{Top: simulated proton image, obtained from particle tracing
simulations of $12~\mbox{MeV}$ protons [as in the data in
Fig.~\ref{channel}~(d)] in the electric field pattern ${\bf
E}(r,z,t)$ given by the 1D particle simulations based on the
ponderomotive, electrostatic model. Bottom : profiles of electric
field $E_r$ and ion density $n_i$, and phase space distributions of
ions $f_i(r,p_r)$ and electrons $f_e(r,p_r)$ from the 1D simulation,
at various times ($t=0$ refers to the laser pulse peak). The
simulation parameters were $a_0=2.7$, $r_0=7.5~\um$, and initial
electron density $n_e=1 \times 10^{19}~\cm3$. \label{PIC1D}}
\end{figure}

Fig.~\ref{PIC1D} shows the electric field ${E}_r(r,t)$ and the ion
density $n_i(r,t)$ obtained from a simulation with $r_0=7.5\lambda$,
$a_0=2.7$ and initial density $0.01n_c$. The initial depletion of
electrons and the later formation of an ambipolar electric field
front are clearly evident. In order to achieve a direct comparison
with the experimental data, a 3D particle tracing simulation,
employing the PTRACE code \cite{romagnaniPRL05}, was carried out to
obtain the proton images for an electric field given by ${\bf
E}(x,r,t)=\hat{\bf r}{E_r}(r,t-x/c)$. The experimental proton source
characteristics (e.g. spectrum and divergence), the detector
configuration and dose response were taken into account. A simulated
proton image, reproducing well the main features observed in the
experiment, is shown in Fig.~\ref{PIC1D}(c). The tip of the channel front is
located $0.75~\psec$ ahead of the pulse peak, in fair agreement with
the previous estimate based on the data.

An essential description theoretical description of the dynamics
observed in the 1D simulations can be given as follows (a detailed
description is reported in Ref.\cite{macchiXXX}). In the first
stage, $F_r$ pushes part of the electrons outwards, quickly creating
a positively charged channel along the axis and a radial ES field
which holds the electrons back, balancing almost exactly the PF,
i.e. $eE_r \simeq F_r$; thus, the electrons are in a
quasi--equilibrium state, and no significant electron heating
occurs. Meanwhile, the force $ZeE_r \simeq ZF_r$ accelerates the
ions producing a depression in $n_i$ around the axis [see
Fig.\ref{PIC1D} (b)]; at the end of the pulse ($t \simeq
2~\mbox{ps}$) we find $E_r\simeq 0$ [see Fig.\ref{PIC1D} (a)],
indicating that the ions have reached the electrons and restored the
local charge neutrality. However, the ions retain the velocity
acquired during the acceleration stage. For $r>r_{max}$, where
$r_{max}$ is the position of the PF maximum, the force on the ions,
and thus the ion final velocity, decrease with $r$; as a
consequence, the ions starting at a position $r_i(0)>r_{max}$ are
ballistically focused towards a narrow region at the edge of the
intensity profile, and reach approximately the same position ($r
\simeq 15~\mu\mbox{m}$) at the same time ($t \simeq 3~\mbox{ps}$,
i.e. $2~\mbox{ps}$ after the peak of the laser pulse); here they
pile up producing a very sharp peak of $n_i$. Correspondingly, the
ion phase space shows that the fastest ions (with energies $\simeq
400~\mbox{keV}$, of the order of the time-averaged ponderomotive
potential) overturn the slowest ones and hydrodynamical breaking of
the ion fluid occurs. Using a simple model \cite{macchiXXX}, the
``breaking'' time and position can be estimated to be $\tau_b \simeq
t_p
+({\pi}/{2\sqrt{2}})e^{3/4}\sqrt{({A}/{Z})({m_p}/{m_e})}({r_0}/{a_0
c})$ (where $t_p=1~\psec$ is the time at which the pulse has maximum
amplitude) and $r_b \simeq ({3}/{2})^{3/2}r_0$, yielding $\tau_b-t_p
\simeq 1.3~\psec$ and $r_b \simeq 14~\um$ for the simulation in
Fig.\ref{PIC1D}, in good agreement with the numerical results. As
inferred from the ion phase space plot at $t=4.7~\psec$, a few ions
acquire negative velocity after ``breaking''; they return toward the
axis and lead to the formation of a local density maximum at $r=0$
after $t\simeq 15~\psec$.

The electron phase space shows that at breaking the electrons are
strongly heated around the ion density peak, generating an ``hot''
electron population with a ``temperature'' $T_h \simeq
13~\mbox{keV}$ and a density $n_h \simeq 4 \times 10^{19}~\cm3$,
corresponding to a local Debye length is $\lambda_D=(T_h/4\pi n_h
e^2)^{1/2} \simeq 0.13~\um$. A modeling of the sheath field thus
generated around the density spike \cite{macchiXXX} (whose thickness
$d \simeq 0.1~\um$ is less than both $\lambda_D$ and the sheath
width $L$) yields a peak field $E_s \simeq 2\pi e n_h d \simeq 6
\times 10^{10}~\Vm$ and a sheath width $L \simeq 4\lambda_D^2/d
\simeq 0.7~\um$, consistently with the simulation results. The
ambipolar field at the ``breaking'' location can be thus be
interpreted as the sheath field resulting from the local electron
heating.

\begin{figure}
\begin{center}
\includegraphics[width=0.8\textwidth]{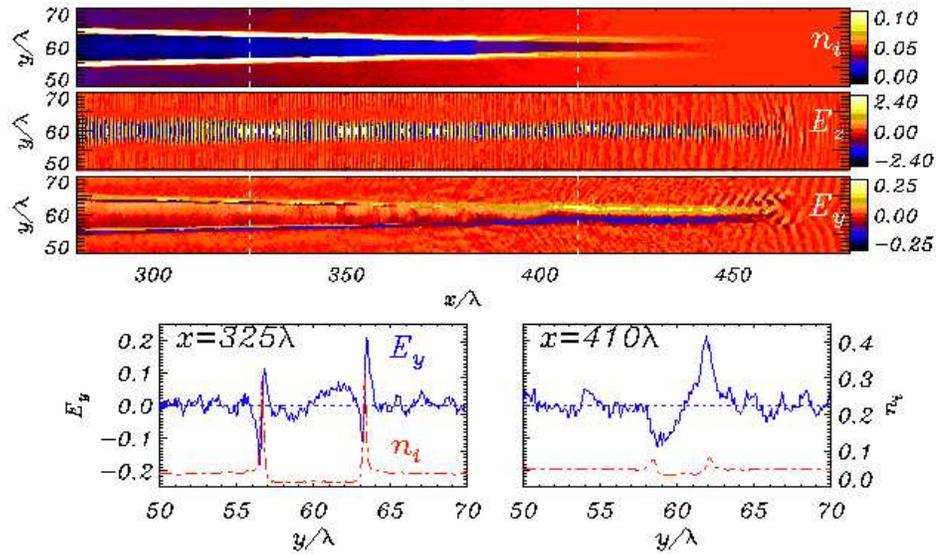}
\end{center}
\caption{Top frame: ion density ($n_{i}$) and electric field
components ($E_{z}$ and $E_{y}$) at $t=2.0~\psec$. Bottom frame:
lineout of $E_{y}$ (blue) and $n_{i}$ (red) along the $y$-axis at
two different $x$-positions, showing the transition in the radial
field pattern. The density is normalized to $n_c=10^{21}~\cm3$ and
the fields to $m_e\omega c/e=3.2 \times 10^{11}~\mbox{V/m}$. The
initial density profile reproduces the experimental one in
Fig.\ref{expt_setup} (b). The laser pulse propagates from left to
right along the $x$ axis and has $1~\psec$ duration. The peak
amplitude is $a_0=2$ in dimensionless units. The laser pulse is
$s$-polarized (i.e. the polarization is perpendicular to the
simulation plane). In this configuration $E_z$ is representative of
the amplitude of the propagating EM pulse and $E_y$ is generated by
the space-charge displacement. The observed increase in $E_z$ by a
factor of $\sim 1.2$ with respect to the peak value in vacuum is due
to self-focusing. \label{fig_EM1}}
\end{figure}

The good agreement of the simulated image with the experimental ones
indicates that the 1D ponderomotive, electrostatic model contains
the essential physics of self-channeling and related electric field
dynamics, despite the exclusion of nonlinear pulse evolution due,
e.g., to self-focusing, or of the plasma inhomogeneity. To address
these issues we performed  electromagnetic particle-in-cell
simulations of the laser-plasma interaction. The simulation were
two-dimensional (2D) in planar geometry - a fully 3D simulation with
spatial and temporal scales close to the experimental ones and
adequate numerical resolutions is way beyond present days
computational power. The 2D results can be considered to be
qualitative since, for instance, the diffraction length of the laser
beam or the scaling with distance of the electrostatic field are
different in 3D. Despite of these limitations, the main features of
the channel observed in the experimental data are qualitatively
reproduced in 2D simulations for a range of parameters close to the
experiment. In the simulation of Fig.~\ref{fig_EM1}, the laser pulse
has a Gaussian intensity profile both in space and time, with peak
dimensionless amplitude $a_0=2$, radius $r_0=4\lambda$ and duration
$\tau_0=300\lambda/c$ where $\lambda$ is the laser wavelength. For
$\lambda=1~\um$ the pulse duration and intensity correspond to
$1~\psec$ and $5.5 \times 10^{18}~\Wcm2$, respectively. The
charge-to-mass ratio of ions is $Z/A=1/2$. The electron density
grows linearly along the $x$-axis from zero to the peak value
$n_0=0.1n_c$ (where $n_c$ is the critical density and
$n_c=10^{21}~\cm3$ for $\lambda=1~\um$) over a length of
$400\lambda$,  and then remains uniform for $200\lambda$. A
$6500\times1200$ numerical grid, a spatial resolution $\Delta
x=\Delta y=\lambda/10$ and $16$ particles per cell for both
electrons and ions were used.

Fig.~\ref{fig_EM1} shows the ion density ($n_i$) and the components
$E_y$ and $E_z$ of the electric field at the time
$t=600\lambda/c\simeq 2.0~\psec$. In this simulation the laser pulse
is $s$-polarized, i.e. the polarization is along the $z$ axis,
perpendicular to the simulation plane. Thus, in Fig.~\ref{fig_EM1}
$E_z$  is representative of the amplitude of the propagating EM
pulse, while $E_y$ is generated by the space-charge displacement.
Simulations performed for the case of $p$-polarization showed no
substantial differences in the electrostatic field pattern. The
simulation clearly shows the formation of an electron-depleted
channel, resulting in an outwardly directed radial space-charge
electric field whose peak value is $6.7 \times 10^{10}~\mbox{V/m}$
(see the lineout at $x=410\lambda$ in Fig.\ref{fig_EM1}). In the
region behind the peak of the pulse, two narrow ambipolar fronts
(one on either side of the propagation axis) are observed. The
ambipolar fields have peak values of $\simeq 6 \times
10^{10}~\mbox{V/m}$ (see the lineout at $x=325\lambda$ in
Fig.\ref{fig_EM1}). As shown above, such a radial electric field
profile produces a pattern in the proton images similar to that
observed in region (III) of Fig.~\ref{channel}.

\section{Conclusion}

We have reported the first direct experimental study of the electric
field dynamics in a charge-displacement channel produced by the
interaction of a high intensity laser pulse with an underdense
plasma. The field profiles  observed clearly identify different
stages of the channel evolution: the electron depletion near the
axis due to the ponderomotive force, and the following ion
acceleration causing a field inversion along the radius. The
features observed are reproduced and interpreted by means of 1D
electrostatic and 2D electromagnetic PIC simulations, followed by a
reconstruction of the proton images employing a 3D particle tracing
code.

\section*{Acknowledgements}

This work has been supported by an EPSRC grant, Royal Society Joint
Project and Short Visit Grants, British-Council-MURST-CRUI, TR18 and
GRK1203 networks, and MIUR (Italy) via a PRIN project. Part of the
simulations were performed at CINECA (Bologna, Italy) sponsored by
the INFM super-computing initiative. We acknowledge useful
discussions with F.~Cornolti and F.~Pegoraro and the support of
J.~Fuchs and the staffs at the Central Laser Facility, RAL (UK).

\end{document}